\documentclass[sigconf]{acmart}
\AtBeginDocument{
  }

\copyrightyear{2026}
\acmYear{2026}
\setcopyright{cc}
\setcctype{by}
\acmConference[KDD '26]{Proceedings of the 32nd ACM SIGKDD Conference on Knowledge Discovery and Data Mining V.1}{August 09--13, 2026}{Jeju Island, Republic of Korea}
\acmBooktitle{Proceedings of the 32nd ACM SIGKDD Conference on Knowledge Discovery and Data Mining V.1 (KDD '26), August 09--13, 2026, Jeju Island, Republic of Korea}
\acmPrice{}
\acmDOI{10.1145/3770854.3783916}
\acmISBN{979-8-4007-2258-5/2026/08}
\usepackage{multirow}
\usepackage{stfloats}
\usepackage{balance}

\begin{document}


\title{Scaling Recommender Transformers to One Billion Parameters}

\author{Kirill Khrylchenko}
\authornote{Work done while at Yandex.}
\orcid{0009-0007-3640-8795}
\affiliation{%
  \institution{HSE University}
  \city{Moscow}
  \country{Russia}
}
\email{elightelol@gmail.com}

\author{Artem Matveev}
\orcid{0009-0004-0271-221X}
\affiliation{%
 \institution{Yandex, HSE University}
  \city{Moscow}
  \country{Russia}
 }
\email{matfu21@ya.ru}

\author{Sergei Makeev}
\orcid{0009-0003-5451-6475}
\affiliation{%
  \institution{Yandex}
  \city{Moscow}
  \country{Russia}
}
\email{neuralsrg@gmail.com}

\author{Vladimir Baikalov}
\authornotemark[1]
\orcid{0009-0009-4864-2305}
\affiliation{%
  \institution{ITMO University}
  \city{Saint-Petersburg}
  \country{Russia}
}
\email{nonameuntitled159@gmail.com}

\renewcommand{\shortauthors}{Kirill Khrylchenko, Artem Matveev, Sergei Makeev, and Vladimir Baikalov}

\begin{abstract}
While large transformer models have been successfully used in many real-world applications such as natural language processing, computer vision, and speech processing, scaling transformers for recommender systems remains a challenging problem. Recently, the Generative Recommenders framework was proposed as a way to scale beyond typical Deep Learning Recommendation Models (DLRMs). By reformulating recommendation as a sequential transduction task, it improves scaling properties in terms of compute. Nevertheless, the largest encoder configuration reported by the HSTU authors is only 176 million parameters --- far smaller than the hundreds of billions (or even trillions) that are now common in language models.

In this work, we present a recipe for training large transformer recommenders with up to one billion parameters. We show that autoregressive learning on user histories naturally decomposes into two subtasks, feedback prediction and next-item prediction, and demonstrate that this decomposition scales effectively across a wide range of transformer sizes. Furthermore, we report a successful deployment on a large-scale music platform serving millions of users. In online A/B tests, the proposed model increases total listening time by +2.26\% and raises the likelihood of user likes by +6.37\%, constituting (to our knowledge) the largest improvement in recommendation quality reported for any deep learning-based system in the platform's history.

\end{abstract}




\maketitle

\section{Introduction}
Recommender systems are an essential part of our daily lives. They help us find inspiration through relevant tracks, pins, and videos; connect with other people and stay informed about the world; and offer a wide range of options for our everyday needs. They also help creators, artists, and vendors find their audiences.

To determine which content to show, recommender systems use machine learning to predict user feedback --- such as whether the user will ignore, like, or dislike a given item. This task poses many challenges: the item catalog is vast and constantly changing, user preferences are dynamic, and user-item interactions are sparse. In such a scenario, complex methods like deep learning become essential.

Deep learning has proven to be a powerful approach for problems involving large amounts of unstructured data, such as text\,\cite{openai} or images\,\cite{alexnet}. While traditional methods rely heavily on manual feature engineering and domain knowledge, neural networks can automatically learn complex patterns from raw inputs. Moreover, the scaling hypothesis\,\cite{openai} claims that this capability grows significantly with increases in training dataset size and model capacity. Over the past decade, scaling has played a pivotal role in fields like computer vision, natural language processing, and speech processing.

Because of extremely large item catalogs and strict latency constraints, the recommendation task is typically approached in multiple stages. The first stage, retrieval, uses lightweight models to filter possible options from the entire catalog. The second stage, ranking, applies more complex architectures to the reduced set of candidates.

There are two predominant neural network archetypes for recommender systems:  
\begin{itemize}
\item \textbf{Early fusion rankers} place emphasis on \emph{feature interactions} across user, item, and user-item features. They typically include an embedding layer\,\cite{unified_embeddings, wide&deep, dlrm}, followed by feature interaction layers\,\cite{dcn-v2} and a feedforward network. Multi-task learning is often employed to predict multiple components of user feedback. Submodules are used to build user and item embeddings, which can be trained either separately as upstream models (e.g., SUM\,\cite{sum}) or jointly with a downstream model (e.g., TransAct\,\cite{transact}, DIN\,\cite{din}). Because these rankers rely on early fusion, they are impractical for retrieval.
\item \textbf{Sequential recommenders} treat users as sequences of events and are usually trained to predict future interactions (e.g., SASRec\,\cite{sasrec}, PinnerFormer\,\cite{pinnerformer}). Due to the linear structure of the decoder layer that produces the next-item probability distribution, these models can be transformed into two-tower architectures with separate user and item encoders, enabling approximate nearest neighbor search\,\cite{hnsw} at the retrieval stage. Moreover, they can also serve as powerful submodules in downstream rankers\,\cite{pinnerformer}.
\end{itemize}
Similarly to other deep learning domains, we aim to leverage the scaling hypothesis in recommender systems. There are four primary scaling options:
\begin{itemize}
\item \textbf{Embeddings.} Recommender models include numerous categorical features with cardinalities varying from 2 to billions\,\cite{unified_embeddings}. Consequently, increasing embedding dimensions for trainable ID-based embeddings rapidly leads to very large embedding matrices. Unfortunately, large embedding layers are crucial for achieving good performance because they represent an information bottleneck between the raw data and the model (see the golf analogy\,\cite{unified_embeddings}). Meta's embedding tables, for instance, range from 675B\,\cite{wukong} to 13T\,\cite{codesign} parameters, while Google has reported roughly a billion parameters since YouTubeDNN\,\cite{youtubednn}. Even Pinterest, a long-time proponent of inductive PinSage embeddings\,\cite{pinsage}, shifted toward large ID-based embedding matrices in their latest work\,\cite{pinterest_id}.

\item \textbf{Context length.} In modern ranking systems, an extensive amount of feature engineering has gone into increasing the context length. The number of features in these systems ranges from hundreds\,\cite{jd, wukong, sum} to thousands\,\cite{hstu}. Meanwhile, user history for sequential recommenders remains relatively short: the gold standard is around 100 interactions\,\cite{transact, youtubednn, facebooknxt, lirank} or 256\,\cite{pinnerformer, kuaishou}.

\item \textbf{Training dataset size.} As \citet{hstu} note, recommender systems can produce training data at a remarkable rate, generating the equivalent of hundreds of GPT-3-scale\,\cite{gpt3} datasets per day. Industry standards have long involved billions of samples: 2B\,\cite{din}, 2.1B\,\cite{kuaishou_ranking}, 3B\,\cite{transact}, 60B\,\cite{codesign}, over 100B\,\cite{dcn-v2, youtubednn}, 146B\,\cite{wukong}, and 500B\,\cite{wide&deep}.

\item \textbf{Encoder capacity.} In early fusion rankers, Google reports dense parts containing between 1M\,\cite{youtube2} and 68M\,\cite{youtube3} parameters in simplified versions of its production models. For sequential recommenders, up to two transformer layers are commonly used\,\cite{transact, pinnerformer, facebooknxt}, with four to five layers being a rarity (e.g., Kuaishou~\cite{kuaishou}). The hidden size typically remains in the low hundreds\,\cite{transact, pinnerformer}, resulting in a few million parameters at most.

\end{itemize}

Although embedding matrices and training datasets are already extremely large, context length and encoder capacity remain underexplored in terms of scaling. We hypothesize that the scaling potential of early fusion rankers is constrained by specialized network layers, such as DCN-v2\,\cite{dcn-v2} or MaskNet\,\cite{masknet}, which impose a strong inductive bias. In contrast, sequential recommenders do not share these structural limitations, yet the encoders reported by both industry and academia remain relatively small.

A significant step forward in scaling research is Meta's Generative Recommenders (GR) framework\,\cite{hstu}. The authors bridge the gap between early fusion rankers and sequential recommenders with a unified generative approach to user modeling. They train a sequential model with a large context (8000 events), a massive dataset (100B samples), and embedding tables on the order of a trillion parameters. Nonetheless, the largest HSTU encoder configuration briefly mentioned in the work (24 layers, 1024-dimensional embeddings) has only 176 million parameters.

A language model with 176 million parameters would perform significantly worse than current largest alternatives. Can we achieve comparable gains when replacing natural language tokens with sequences of user-interaction events? Successfully scaling recommender transformer encoders could result in significant betterment of personalized services, benefiting billions of users worldwide through a more nuanced understanding of user interests. In this work, we focus on scaling recommender transformer encoders.

The main contributions of the paper are as follows:
\begin{itemize}
\item Unlike Monty Python's King Arthur\footnote{\url{https://www.imdb.com/title/tt0071853/}}, we do attain our goal: we scale recommender transformers to a billion parameters and observe a significant improvement in recommendation performance.  
\item We propose a fundamental pre-training task that naturally decomposes into two complementary objectives: user feedback prediction and next-item prediction, while scaling effectively across a wide range of model sizes.
\item We introduce a computationally efficient fine-tuning stage that converts the large transformer encoder into a two-tower architecture, enabling offline inference and providing a powerful ranking feature for downstream models.
\item We deploy a 126M-parameter transformer model with context length 8192 on an industry-leading music platform with millions of users and items. To our knowledge, this deployment yielded the greatest quality improvements among all neural-network-based recommender models at that platform.
\end{itemize}
We dub our approach ARGUS (\textbf{A}uto\textbf{R}egressive \textbf{G}enerative \textbf{U}ser \textbf{S}equential framework).

\section{Related work}

\paragraph{Scaling deep learning}
AlexNet\,\cite{alexnet} was the first major success of deep learning and also a key milestone for scaling, as ImageNet was significantly larger than any previous dataset. \citet{google_scaling_1} improved an ImageNet classifier by pre-training on the large-scale but noisy JFT dataset, demonstrating that performance scaled logarithmically with dataset size. \citet{baidu} validated scaling laws across machine translation, language modeling, image processing, and speech recognition. \citet{facebook_scaling_1} used large-scale weakly supervised data from Instagram to pre-train an ImageNet classifier. \citet{facebook_scaling_2} showed that limited model capacity reduces the gains from data scaling; they also highlighted the importance of self-supervised learning and the need to identify appropriate pre-training tasks. \citet{openai} formulated scaling as a compute allocation problem, exploring trade-offs between model size and dataset size; they concluded that larger models are more sample-efficient and that increasing model size nearly always yields better performance. \citet{chinchilla} showed that, due to suboptimal training (including learning-rate scheduling), the models proposed by \citet{openai} were undertrained. \citet{gpt2} explored language modeling in depth, framing next-token prediction as an extreme multi-task learning problem that enables strong scaling behavior.

\paragraph{Early fusion neural rankers}
Wide\&Deep\,\cite{wide&deep} combines a deep neural network (DNN) with a linear model for ranking, whereas YouTubeDNN\,\cite{youtubednn} discards the linear component entirely. There is substantial research on cross-feature and feature-interaction modeling\,\cite{fm, deepfm, dcn-v1, autoint, dlrm, dcn-v2, hiformer} for early fusion rankers. Such models often include extremely large embedding matrices with up to 12 trillion parameters\,\cite{codesign}, while the dense encoder part typically contains only tens of millions of parameters at most. Another line of research examines scaling of neural rankers. \citet{meta_scaling_1} tested scaling DLRMs\,\cite{dlrm} for CTR prediction and concluded that ``\textit{parameter scaling is out of steam for the model architecture under study, and until a higher-performing model architecture emerges, data scaling is the path forward}''. Wukong\,\cite{wukong} reintroduces the concept of stacked factorization machines and demonstrates scaling up to 10B parameters. However, to our knowledge, subsequent work has not reproduced Wukong's scaling results. Similarly to \citet{ligr}, our own experiments with Wukong did not replicate the reported outcomes.

\paragraph{Sequential modeling for ranking}
YouTubeDNN\,\cite{youtubednn} applies average pooling over the last-watched videos and search queries to form user embeddings, which are then used in both ranking and retrieval. DIN\,\cite{din} employs pointwise target-aware attention on user history. BST\,\cite{bst}, TransAct\,\cite{transact}, and LiRank\,\cite{lirank} incorporate a small target-aware transformer over recent user history as a submodule in the downstream ranker. HSTU\,\cite{hstu} reframes impression-level ranking as a generative task by interleaving actions and items into a single sequence.

\paragraph{Sequential modeling for retrieval}
YouTubeDNN\,\cite{youtubednn} formulates retrieval as the prediction of a user's next video watch. CASER\,\cite{caser}, GRU4Rec\,\cite{gru4rec}, and SASRec\,\cite{sasrec} represent the user as a sequence of positive user-item interactions, training CNNs, RNNs, and transformers, respectively, to predict the next positive interaction. PinnerFormer\,\cite{pinnerformer} additionally includes negative user-item interactions in the user history and trains to predict future positive interactions. HSTU\,\cite{hstu} retrieval adopts a similar approach, predicting the next positive interaction based on the full interaction history. \citet{latent_cross} proposed an RNN-based retrieval model for YouTube, while \citet{topk} framed recommendation as a reinforcement learning problem and applied REINFORCE with off-policy correction to the same model\,\cite{latent_cross}.

\paragraph{Scaling sequential recommenders}
CLUE\,\cite{clue} trains a transformer encoder on user histories from multiple domains with contrastive learning, demonstrating scaling with respect to training data, context length, and model size. \citet{naver} represented the user as a sequence of textual item descriptions and trained a transformer encoder to predict the next item; they also reported benefits from scaling model size. \citet{amazon} trained a transformer on next-event prediction, decomposing events into separate features, and scaled model size up to 85M parameters. \citet{wechat} trained a transformer on next-item prediction, reporting a scaling law even better than in NLP. \citet{Huawei} explored scaling for the HSTU architecture, which we discuss further. However, most of these studies use the well-known \emph{leave-one-out} evaluation scheme without a proper temporal split\,\cite{timesplit1, timesplit2}. While this may be acceptable for small models, omitting a temporal split for large models is problematic due to their memorization capacity. HSTU\,\cite{hstu} addresses this issue by evaluating on a proprietary industrial dataset with an appropriate temporal protocol, and by reporting online metrics. The authors scaled their encoder to an 8k context and 100B samples and introduced a new architecture. Still, the largest encoder mentioned contains just 176M parameters.

\section{Model architecture}
\subsection{Pre-training}
Scaling deep learning is \emph{guaranteed to succeed} if: a model (1) with sufficient capacity is (2) pre-trained on a fundamental task (3) with massive amounts of data. In recommender systems, user feedback generates immense amounts of training data, and transformers seem a natural fit for modeling user-history sequences. But a key question remains: \emph{what should the training task be?} For almost a decade, the industry standard has been next-item prediction, where the model is trained to predict the next positive user-item interaction. Yet, it has not shown clear scaling benefits in practice. In this section, we propose a pre-training objective that scales across a wide range of model sizes.

Consider large language models: despite being trained on noisy, large-scale internet data, they still produce reasonable responses. Prompting such a model often yields an average internet-style answer, which may be suboptimal or only partially accurate. However, modifying the prompt --- for instance, with a prefix such as ``\textit{Let's say you are very knowledgeable}'' --- can shift the output distribution toward objectively better responses\,\cite{gpt3}. This reflects the model's ability to leverage both the prefix context and its internal world knowledge, acquired during pre-training, to refine its outputs. In reinforcement learning terminology, the model improves the \emph{logging policy} it imitates (internet answers) by incorporating its understanding of the environment (world knowledge and abstract patterns). Both imitation of the logging policy and world knowledge accumulation occur during pre-training.

\begin{figure*}[t]
  \centering
  \includegraphics[width=1.\textwidth]{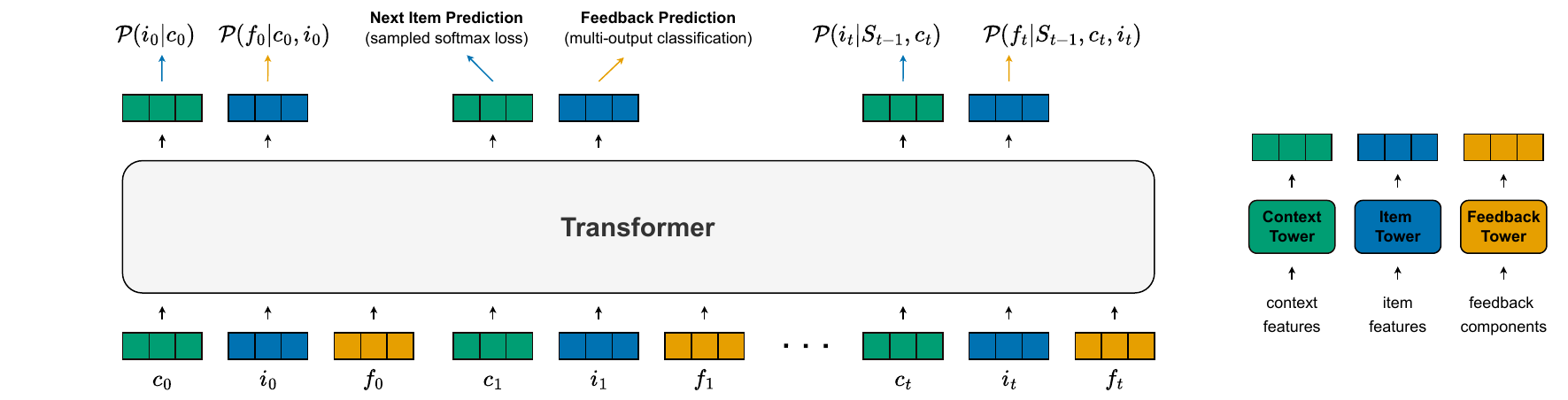}
   \Description{A horizontal diagram illustrates a sequence of triplets $(c_0, i_0, f_0), (c_1, i_1, f_1), \dots,$ along the bottom, where each triplet contains context  $(c_t)$, item $(i_t)$, and feedback$(f_t)$. A large Transformer block in the center takes these triplets as inputs from left to right. Two parallel "heads" appear at the top: one for next-item prediction, the other for feedback prediction. On the far right are three small towers labeled "Context", "Item", and "Feedback", representing the separate embedding spaces. The figure shows how a use's context-item-feedback history is simultaneously used for both next-item and feedback modeling within a single Transformer.}
  \caption{An overview of the pre-training architecture, which takes the user's context-item-feedback history as input to a transformer. Two parallel heads produce next-item predictions and feedback predictions, respectively.}
\label{fig:pretrain}
\end{figure*}

\begin{table}[t]
\caption{Notation}
\begin{tabular}{l|l}
     Notation & Description \\
     \hline
     $c_t \in \mathcal{C}$ & Context (e.g., surface, device, location) \\
     $i_t \in \mathcal{I}$ & Item (e.g., music track) \\
     $f_t \in \mathcal{F}$ & Feedback (e.g., like, skip) \\
     $(c_t, i_t, f_t)$ & $t$-th user-item interaction \\
     $S_T := \{ (c_t, i_t, f_t) \}_{t=1}^T$ & Historical user interaction sequence \\
     $h_t^c, h_t^i$ & Encoder hidden states for $c_t$ and $i_t$ \\
     $h_t$ & Hidden state of the simplified model \\
     $\theta \in \mathbb{R}^{D}$ & Trainable model parameters \\
     \hline
\end{tabular}
\end{table}

Motivated by our analogy to large language models, we conceptually reframe recommendation as a reinforcement learning problem, similarly to \citet{topk}:
\begin{itemize}
\item The recommender system is an \textbf{agent}.  
\item $\mathcal{A} := \mathcal{I}$. The \textbf{action space} corresponds to items recommended to users; in the simplest case, an action consists of recommending a single item.
\item User interests, browsing habits, and interaction patterns define the \textbf{environment}, which exists independently of any single recommender.
\item $\mathcal{S} := \bigcup_{T=1}^{\infty} \left( \mathcal{C} \times \mathcal{I} \times \mathcal{A} \right)^T$. The \textbf{state} captures user history and naturally satisfies the Markov property.
\item $\pi_\theta(a \mid s) := \pi_\theta(\text{item} \mid \text{history}, \text{context})$. The model's \textbf{policy} maps user states to item distributions.
\end{itemize}

Within this formulation, the production recommender system that generated the training data acts as a \textbf{logging policy} --- the behavior we can directly imitate. Concurrently, user behavior provides \textbf{world knowledge}: insights about preferences that are not tied to any particular system, and are visible through user feedback and organic\footnote{Organic user events are those not influenced by explicit system recommendations (e.g., search activity).} transitions. Drawing on our language model analogy, we propose a pre-training task with dual objectives: (1) learning to imitate prior recommender systems and (2) learning user preferences from feedback and organic navigation patterns.

Note that our training is fully offline and supervised; we do not perform online reinforcement learning. We use RL terminology as a conceptual lens: next-item prediction corresponds to policy imitation, while feedback prediction encourages learning user response regularities that can be interpreted as environment modeling.

\paragraph{Next-item prediction.} 
In our approach, the model sees a sequence of context-item-feedback triplets $(c_t, i_t, f_t)$, where $c_t$ includes contextual features such as the surface\footnote{Surface refers to the interface or platform where a user is exposed to an item, e.g., organic discovery (browsing, search) versus algorithmic recommendations.}, $i_t$ is the item that appears, and $f_t$ is user feedback. The next-item prediction task is: $$\mathcal{P} (\text{item}=i_t \mid \text{history}=S_{t-1}, \text{context}=c_t).$$

We train on all item interactions, including non-positive feedback. Crucially, $c_t$ identifies the interaction surface, which indicates whether the exposure is recommended or organic and, for recommended impressions\footnote{An impression is an item explicitly shown to the user by the recommendation system.}, often identifies the serving policy specific to that surface. Conditioning on $c_t$ therefore lets a single model separate these interaction types: it can reproduce surface-specific recommendation behavior while also learning organic navigation patterns that reflect user preferences.

To optimize this task, we use logQ-corrected sampled softmax\,\cite{logq} with mixed negative sampling\,\cite{mns}:
\begin{equation*}
    \mathcal{L}_{\text{NIP}}(S_{t-1}, c_t, i_t; \theta) = -\log \frac{e^{f(h_t^c, i_t)}}{e^{f(h_t^c, i_t)} + \sum\limits_{n \in N} e^{f(h_t^c, n) - \log Q(n)}},
\end{equation*}
where:
\begin{itemize}
    \item \(f(h_t^c, i_t) := \cos (h_t^c, i_t) / \exp(\tau) \) is a similarity measure between the context-aware embedding \(h_t^c\) and the item embedding \(i_t\),
    \item $N = N_{\text{in-batch}} \cup N_{\text{uni}}$ consists of $|N_{\text{in-batch}}|=8192$ in-batch negatives and $|N_{\text{uni}}|=8192$ uniformly sampled items,
    \item $Q(n)$ is an estimate of the negative sampling probability under our mixed sampler (maintained with a count-min sketch as in~\cite{logq}),
    \item $\exp(\tau)$ is a learnable temperature, clipped to $[0.01, 100]$.
\end{itemize}

\paragraph{Feedback prediction.} While next-item prediction captures logging policy behavior and organic user behavior patterns, modeling actual user preferences requires an additional feedback-prediction component:
\[\mathcal{P}(\text{feedback} = f_t \mid \text{history} = S_{t-1},\text{context} = c_t,\text{item} = i_t).\]
In practice, feedback is often multivariate (e.g., skip/like, listening duration) and can be decomposed into K factors: \(\mathcal{F} = \prod_{k=1}^K \mathcal{F}_k\). To simplify, we assume these components to be conditionally independent given the state:
$\mathcal{P}_\theta(f_t \mid h_t^i)
= \prod_{k=1}^K 
  \mathcal{P}_\theta(f_t^k \mid h_t^i).
$ This turns feedback prediction into a multi-task learning problem. The overall loss is:
\begin{align*}
\mathcal{L}_{\text{FP}} &= -\log \prod\limits_{k=1}^K \mathcal{P}_\theta(f_t^k \mid h_t^i) = \sum\limits_{k=1}^K \text{CrossEntropyLoss}(f_t^k, l_t^k),
\end{align*}
where $l_t^k$ is the logit for the $k$-th feedback dimension. This objective complements next-item prediction by explicitly modeling how users react to items.

\begin{figure}[t]
\hspace*{-1cm}
  \centering
  \includegraphics[width=0.4\textwidth]{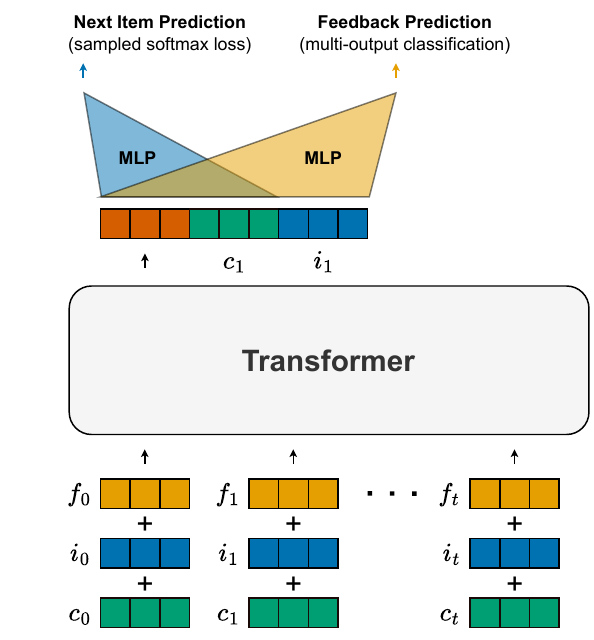}
  \Description{
  A simplified pre-training architecture that merges each context-item-feedback triplet into a single embedding vector. Along the bottom, triplets  $(c_0, i_0, f_0), (c_1, i_1, f_1), \dots,$ appear as stacked colored blocks (for example, context in red, item in blue, feedback in orange). These per-interaction embeddings feed into a single Transformer. Above the Transformer is a schematic of two MLP heads: one producing next-item predictions (sampled softmax), the other producing feedback predictions (multi-output classification). The visual emphasizes how reducing the input sequence length (from $3n$ to $n$) improves efficiency but removes the option to treat context, item, and feedback as separate tokens.}
  \caption{Simplified pre-training architecture that merges each context-item-feedback triplet into a single embedding}
    \label{fig:simplified}
\end{figure}
We combine these two objectives into our final pre-training loss, $\mathcal{L}_{\text{pre-train}} = \mathcal{L}_{\text{NIP}} + \mathcal{L}_{\text{FP}}$, as illustrated in Figure~\ref{fig:pretrain}.

\paragraph{Differences with SASRec.} Traditional sequential models such as SASRec\,\cite{sasrec} also rely on next-item prediction, but typically consider only positive interactions. This conflates the likelihood of an item being shown with the likelihood of it receiving positive feedback, and it overlooks many neutral interactions that are essential for modeling real-world behavior. In contrast, our model incorporates both impressions resulting from system recommendations (regardless of user response) and user-driven (organic) events, and also explicitly models user feedback.

\paragraph{Simplified architecture.}
Representing each user-item interaction as a triplet $(c_t, i_t, f_t)$ leads to a sequence of length $3n$ for $n$ interactions, which becomes computationally expensive for long user histories. To reduce complexity, we merge each triplet into a single interaction embedding. Concretely, the encoder now outputs one hidden state $h_t$ per interaction instead of three (see Figure~\ref{fig:simplified}).

This compression introduces trade-offs in how context and item information are represented and reused across tasks. For next-item prediction, we can no longer obtain a fully context-aware hidden state $h_t^c$. Instead, we approximate it by concatenating the previous hidden state $h_{t-1}$ with the current context embedding $c_t$, followed by an MLP projection:
\begin{equation*}
    \hat{h}^c_t = \text{MLP}(\text{Concat}(h_{t-1}, c_t)).
\end{equation*}
Similarly, for feedback prediction we lose direct access to the target item (target awareness\,\cite{din}). We approximate the target-aware state as:
\begin{equation*}
    \hat{h}^i_t = \text{MLP}(\text{Concat}(h_{t-1}, c_t, i_t)).
\end{equation*}

\subsection{Fine-tuning}\label{sec:fine-tuning}
There are multiple ways to adapt our pre-trained model to downstream tasks. In this work, we focus on fine-tuning and leave other adaptation methods for future work.

Our primary downstream objective is item reranking. A widely adopted training approach is impression-aware pointwise training\,\cite{hstu}, defined as:
\begin{equation*}
    \mathcal{L}_{\text{ranking}} (u, i, f) = \text{CrossEntropyLoss}(f, \phi_\theta(u,i)),
\end{equation*}
where user $u$ is shown item $i$ and provides feedback $f$ (e.g., a like), and $\phi_\theta(u,i)$ is a ranking model that incorporates user, item, and user-item features\,\cite{dlrm}.

In practice, the pointwise loss can be replaced by pairwise or listwise alternatives depending on the application. In our production setting, we use a pairwise ranking objective, which is discussed in more detail in Section~\ref{sec:dataset}.

\begin{figure}[t]
\hspace*{-1.5cm}
  \centering
  \includegraphics[width=0.58\textwidth]{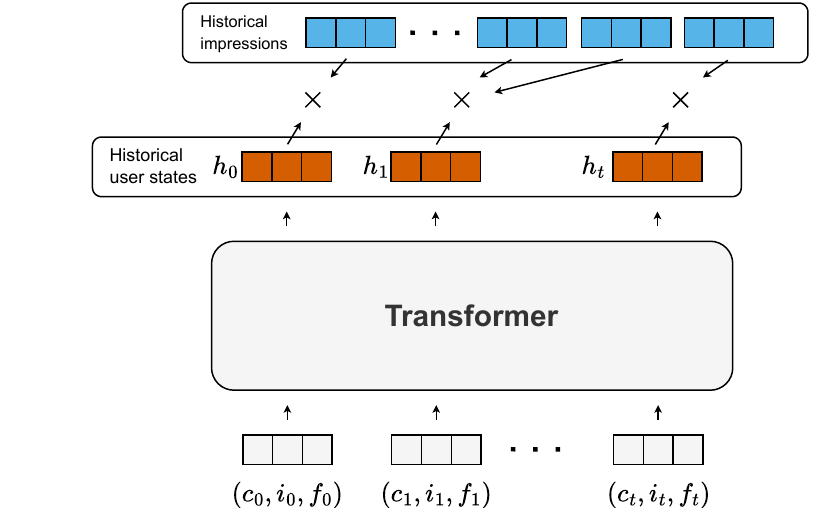}
  \Description{
  Impression-aware, one-pass, two-tower fine-tuning with a single causal-encoding Transformer. The lower row shows user interaction triplets $(c_0, i_0, f_0), (c_1, i_1, f_1), \dots, (c_t, i_t, f_t)$ entering the Transformer. At the top, the Transformer outputs a sequence of user representations $h_0, h_1, \dots, h_t$. A separate row labeled "Historical Impressions" displays recommended items aligned by timestamp or simulated latency; lines or "$\times$" marks indicate how these impressions map to the corresponding user states. Finally, each impression embedding is scored against its matched user state via a dot product (e.g., $h_t \dot i_t$). This score drives a ranking loss for downstream recommendation tasks.
  }
  \caption{Impression-aware, one-pass, two-tower fine-tuning with a single causal-encoding transformer}
    \label{fig:fine-tuning}
\end{figure}

Conceptually, this ranking objective aligns with the pre-training goal of modeling feedback $\mathcal{P}(\text{feedback} \mid \text{history}, \text{context}, \text{item})$. However, practical considerations and task-specific constraints motivate a separate fine-tuning procedure:
\begin{itemize}
\item \textbf{Domain shift.} Our pre-training domain is broader than the final task domain: the model is initially exposed to feedback across all contexts (including purely organic ones), whereas the final ranking only needs to predict feedback for recommended items. During fine-tuning, we restrict training to items with recommendation-based contexts.
\item \textbf{Daily inference.} Pre-training assumes zero delay for data delivery and relies on resource-intensive real-time inference. Due to efficiency concerns, we prefer to offload computation to offline batch workloads. A common approach\,\cite{pinnerformer} is to train a two-tower model with simulated data delivery delay: we compute user and item embeddings once per day, store user embeddings in a key-value system, then use dot products at serving time.
\item \textbf{Causal pre-training.} Simulating 24-hour data delivery delay or adopting complex pairwise or listwise ranking losses during pre-training would require more complex attention masking schemes. In contrast, our fine-tuning procedure retains the simple causal masking used during pre-training.
\end{itemize}

Figure~\ref{fig:fine-tuning} illustrates our impression-aware, one-pass, two-tower fine-tuning procedure, designed to address the challenges above. We run a single pass over the user's history through a causal transformer, forming a sequence of user representations. Impressions (recommended items) form a separate sequence and are matched to user hidden states by timestamp and simulated latency. We then compute dot products between each impression's embedding and the corresponding user representation to obtain a ranking score, feeding the result into a suitable ranking loss (e.g., pointwise or listwise):
$$\phi_\theta(u, i_t) = \left\langle h_t, i_t \right\rangle,$$
where $i_t$ denotes the embedding of an impressed item and $h_t$ is the timestamp-aligned user representation.

We present this configuration as a coherent scaling recipe rather than a set of independent heuristics. While individual components like feedback prediction or specific attention mechanisms could be further ablated, our results demonstrate that this specific combination provides a stable and reproducible path to scaling recommender transformers to the billion-parameter regime.

\section{Experiments}
We aim to answer the following research questions:
\begin{itemize}
    \item \textbf{RQ1:} Does ARGUS scale effectively?
    \item \textbf{RQ2:} How does HSTU compare to a standard transformer encoder?
    \item \textbf{RQ3:} Is our two-stage training pipeline necessary? Can we simplify it?
    \item \textbf{RQ4:} Does context length scaling improve recommendation quality in the music domain?
    \item \textbf{RQ5:} How does ARGUS perform on real-world recommendation scenarios?
\end{itemize}
\subsection{Experimental Setup}
In this section, we describe our datasets, baselines, evaluation protocol, and implementation details for the ARGUS framework.

\subsubsection{Datasets}\label{sec:dataset}
Dataset selection was based on the following criteria:
\begin{itemize}
    \item \textbf{Scalability:} The dataset must be sufficiently large to support training and evaluation of models with hundreds of millions or even billions of parameters. Small datasets would either lead to underfitting in a one-epoch regime or overfitting when trained for multiple epochs.
    \item \textbf{Content Coverage:} To learn both next-item and feedback-prediction tasks as defined above, the dataset should include both impressed items and user feedback, as well as various contexts --- recommendation-based and organic ones.
    \item \textbf{Applicability:} The dataset should reflect real-world usage scenarios and be applicable for validating the model in a production environment. Ideally, leading to an A/B test.
\end{itemize}
Since no public dataset met these requirements (to our knowledge)\footnote{Since the completion of this work, the Yambda dataset\,\cite{yambda} has been published, which is quite similar to our production dataset. However, it contains fewer features, and is substantially smaller in scale.}, we constructed our own dataset by sampling one year of activity for tens of millions of users from our music-streaming platform. This dataset contains \emph{over 300B user-item interactions} across millions of items, including ItemID, ArtistID, and other item features. It covers implicit signals such as listening duration and skips, explicit feedback such as likes, and contextual features including surface type, device, and recommendation settings.

For pre-training, we split each user's history into fixed-length chunks. Each chunk is treated as a separate training sample. For fine-tuning, we use a pairwise logistic loss, forming impression pairs from temporally adjacent interactions with different feedback outcomes.

\subsubsection{Baselines}
To evaluate the effect of model scaling and two-stage training, we primarily compare models within the same architecture family. We fix our modeling family to a standard transformer applied over user interaction histories, which serves as a general and representative baseline for studying scaling behavior. In addition, we compare against a drop-in encoder replacement by swapping the transformer for HSTU\,\cite{hstu}, whose authors argue for improved scaling compared to transformers on recommendation tasks.

Our main benchmark for downstream utility is against our \textbf{production ranking model}: a gradient-boosted decision tree ensemble trained with pairwise logistic loss on one thousand features. These features include standard heuristics (counters, ratios), handcrafted signals, and outputs from earlier generations of \textbf{transformer-based two-tower models}. We describe these prior models in more detail in section~\ref{sec:real-world}.

For downstream evaluation, we do not compare against traditional sequential recommenders such as SASRec\,\cite{sasrec}. Such models are typically optimized for next-item prediction and are primarily used in retrieval-style settings, rather than impression-level ranking; consequently, they would be a poor fit and would substantially underperform in our ranking setup. We also do not compare against target-aware ranking formulations of HSTU\,\cite{hstu}: they rely on real-time data and explicit target awareness (conditioning on the candidate item), which is computationally much more expensive than our two-tower setup with daily processing of user sequences and offline inference.
    
\subsubsection{Evaluation Metrics}\label{sec:eval}
We evaluate models using a \emph{global temporal split}, holding out the week following the training period as a test set. Unlike random splits often used in scaling studies, strict temporal evaluation is crucial for large-scale models. It prevents performance overestimation caused by data leakage (memorization of future interactions) and inherently tests the model’s robustness to distribution drift.

To assess the pre-trained model, we focus on two core capabilities: next-item prediction and user feedback modeling:
\begin{itemize}
\item \textbf{Feedback prediction.} We measure normalized entropy (as in\,\cite{facebook_practical}) with respect to a baseline feedback distribution estimated from empirical frequencies.
\item \textbf{Next-item prediction.} Normalized entropy is computed relative to a unigram item distribution computed from user-item interactions, using the same sampled softmax setup (8192 uniform and 8192 in-batch negatives) as in training.
\end{itemize}

After pre-training, we evaluate the model's ability to rank impressions in two scenarios:  
\begin{enumerate}
    \item \textbf{Standalone ranking:} the model directly ranks impressions.   
    \item \textbf{Feature integration:} the model's output is provided as an additional feature to our production ranker.
\end{enumerate}
We measure pairwise accuracy, which compares the model's ordering against actual user preferences. Formally, let $\Omega$  be the set of temporally adjacent impression pairs $i_1, i_2$ where $i_1$ received more positive feedback than $i_2$ (e.g., skipped track is worse than a track with completion rate above a certain threshold, which is worse than an explicit like). Then
\[
\mathrm{PA}(\text{model}) \;=\; 
    \frac{1}{|\Omega|}\sum_{(i_1, i_2)\in \Omega} 
    \begin{cases}
        1, & \text{if } \mathrm{score}(i_1) > \mathrm{score}(i_2),\\
        0.5, & \text{if } \mathrm{score}(i_1) = \mathrm{score}(i_2),\\
        0, & \text{otherwise}.
    \end{cases}
\]
We then define the pair accuracy uplift relative to our production ranker:
\[
\mathrm{PAU}(\text{model}) \;=\; 
    \frac{
        \mathrm{PA}(\text{model}) \;-\; \mathrm{PA}(\text{prod})
    }{
        \mathrm{PA}(\text{prod})
    }
    \times 100\%,
\]
where PA(prod) is the pairwise accuracy of the production ranker. Positive PAU indicates improvement over the baseline, while negative values indicate a performance drop. Unlike standard ROC-AUC, which aggregates over all comparable pairs, our metric restricts comparisons to temporally adjacent impressions within a user sequence.

The negative uplift in standalone ranking (e.g., -4.78\% for Large) should be interpreted in context: our model operates in an offline mode with a 24-hour data lag and uses only user history. In contrast, the production baseline is a complex ensemble that leverages hundreds of real-time features available at serving time. The fact that a single offline transformer can approach this level of performance underscores its representational power.

\paragraph{Implementation Details}
\begin{table}[!t]
\centering
\caption{Hyperparameter overview (pre-train vs. fine-tune)}
\label{tab:hyperparams}
\begin{tabular}{lcc}
\toprule
\textbf{Hyperparameter} & \textbf{Pre-train} & \textbf{Fine-tune} \\
\midrule
Effective batch size & 4096 & 2048 \\
\midrule
Dropout & \multicolumn{2}{c}{0.1} \\
Optimizer & \multicolumn{2}{c}{Adam} \\
Grad norm clipping & \multicolumn{2}{c}{1} \\
Number of warmup steps & \multicolumn{2}{c}{3000} \\
Backbone LR Schedule & \multicolumn{2}{c}{$10^{-5} \to 10^{-4} \to 10^{-4} $} \\
Head LR Schedule & \multicolumn{2}{c}{$10^{-3} \to 10^{-3} \to 10^{-4}$} \\
\bottomrule
\end{tabular}
\end{table}

\begin{table*}[!t]
\centering              
\caption{ARGUS model scaling: from 3.2M to 1B parameters. Results are reported for pre-training and fine-tuning tasks. Parameter counts include encoder weights only.}
\begin{tabular}{lcrccccc}
\toprule
\multicolumn{3}{c}{\centering \textbf{Encoder}} & \multicolumn{3}{c}{\textbf{Pre-train}} & 
\multicolumn{2}{c}{\parbox{2.7cm}{\centering \textbf{Ranking Fine-tune}\\\small (pair accuracy uplift)}} \\
\cmidrule(lr){1-3}
\cmidrule(lr){4-6}
\cmidrule(lr){7-8}
\multirow{2}{*}{\centering \textbf{Model}}
  & \multirow{2}{*}{\centering \textbf{Configuration}}
  & \multirow{2}{*}{\centering \textbf{\#Params}}
& \multicolumn{2}{c}{\parbox{4cm}{\centering \textbf{Feedback Prediction}\\\small (normalized entropy)}}
  & \multirow{2}{*}{\parbox{2.7cm}{\centering \textbf{Next-Item Prediction}\\\small (normalized entropy)}}
  & \multirow{2}{*}{\centering \textbf{Standalone}}
  & \multirow{2}{*}{\centering \textbf{Feature}} \\
\cmidrule(lr){4-5}
\multicolumn{1}{c}{} 
  &
  &
  & \textbf{Consumption}
  & \textbf{Engagement}
  & 
  & 
  & \\
\midrule
    Mini & L4 H256 & 3.2M  & 0.5830 (-0.00\%) & 0.5855 (-0.00\%) & 0.4777 (-0.00\%) & -7.49\% & +1.35\% \\
    Small & L6 H512 & 18.9M & 0.5756 (-1.28\%) & 0.5707 (-2.51\%) & 0.4691 (-3.81\%) & -6.29\% & +1.82\% \\
    Medium & L10 H1024 & 126.0M  & 0.5690 (-2.41\%) & 0.5556 (-5.10\%) & 0.4502 (-7.68\%) & -5.33\% & +2.32\% \\
    \textbf{Large}  & \textbf{L20 H2048} & \textbf{1.007B} & \textbf{0.5631 (-3.43\%)} & \textbf{0.5436 (-7.15\%)} & \textbf{0.4372 (-10.36\%)} & \textbf{-4.78\%} & \textbf{+2.66\%} \\
\midrule
HSTU & L24 H1024 & 176.0M & 0.5684 (-2.51\%) & 0.5553 (-5.15\%)& 0.4488 (-7.99\%) & -5.39\% & +2.34\% \\
\bottomrule
\end{tabular}
\label{table:scaling}
\end{table*}

All experiments use PyTorch 2.x with Distributed Data Parallel (DDP) across 64--256 A100 80GB GPUs. Training duration ranges from 1 day to 1 week, depending on the model size. Following production practice\,\cite{hstu}, we train for a single epoch over the dataset. Unless otherwise specified, all models are trained with the following configuration:
\begin{itemize}
    \item \textbf{ARGUS (simplified)} --- no context-item-feedback interleaving; each interaction is represented as a single embedding
    \item  \textbf{Unified embeddings} for categorical features\,\cite{unified_embeddings} --- we use a 3-way lookup for item ID. The same embedding matrix size is used across all experiments and contains 130M parameters (512k embeddings of size 256). Note that these are additional parameters that are not counted towards encoder size.
    \item \textbf{Absolute trainable positional embeddings}
    \item \textbf{Output embedding size:} 512 for both users and items
    \item \textbf{Sequence length:} 512 (pre-training), 2048 (fine-tuning)
    \item \textbf{Encoder:} medium transformer configuration (L10 H1024)
    \item \textbf{Latency simulation during fine-tuning:} 24-hour delay for matching impressions to user states
\end{itemize}
Table \ref{tab:hyperparams} summarizes the main hyperparameters. We schedule distinct learning rates for the backbone vs. the task-specific heads: backbone parameters use linear warmup followed by a constant rate, while head parameters use a larger learning rate with linear warmup followed by linear decay.

\subsection{Scaling Hypothesis (RQ1)}

\begin{figure}[!t]
  \centering
  \includegraphics[width=\linewidth]{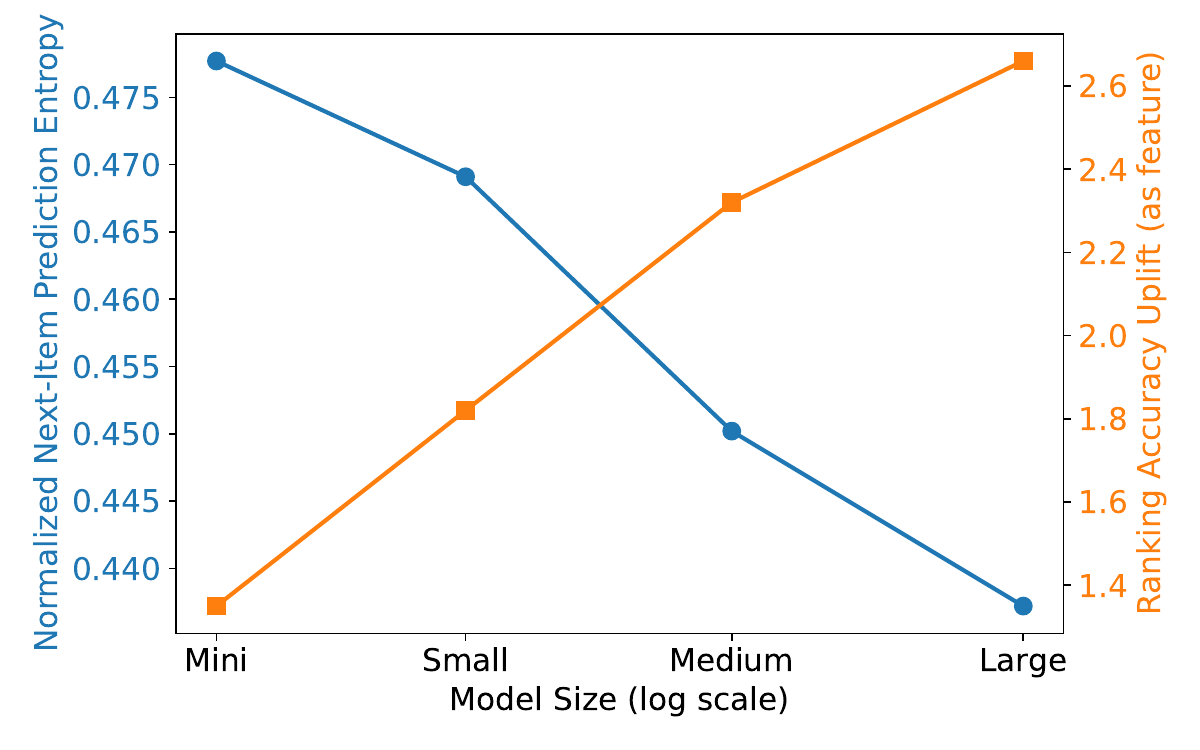}
  \caption{ARGUS model scaling: normalized next-item prediction entropy and ranking accuracy uplift. This plot is derived from Table~\ref{table:scaling}. Note the linear trend on the log scale, suggesting that the scaling laws observed in LLMs hold for our proposed recommender architecture.}
  \Description{A line plot comparing the performance of ARGUS models of increasing size (Mini, Small, Medium, Large) across two metrics: normalized next-item prediction entropy and ranking accuracy uplift (as a feature). The x-axis shows model size on a logarithmic scale, labeled as Mini, Small, Medium, and Large. The y-axis shows metric values. As model size increases, next-item entropy decreases steadily, and ranking accuracy uplift improves. The plot illustrates a clear scaling trend.}
  \label{fig:scaling-graph}
  \vspace{-3mm}
\end{figure}

In Table~\ref{table:scaling}, we compare four versions of ARGUS encoders ranging from 3.2M (Mini) to 1B (Large) parameters. We select model sizes on an approximately logarithmic parameter scale to enable a consistent evaluation of scaling laws.

We evaluate both pre-training performance (via normalized entropy for feedback and next-item prediction) and downstream performance (via pairwise accuracy uplift on temporally adjacent item impressions).

As model size increases, we observe consistent improvements across all metrics. Feedback prediction entropy improves by 3 -- 7\%, next-item entropy drops by over 10\%, and pairwise accuracy uplift increases from +1.35\% (Mini) to +2.66\% (Large). These results, also visualized in Figure~\ref{fig:scaling-graph}, suggest clear scaling trends for transformer-based recommenders.

\subsection{Encoder: Transformers vs. HSTU (RQ2)}
\label{sec:rq2-hstu}
HSTU\,\cite{hstu} is a recently proposed encoder architecture positioned as an alternative to transformers for recommendation. The authors argue that it better captures counting-style signals and report strong results and favorable scaling when replacing a transformer encoder on academic benchmarks. Since our work studies scaling in transformer-based recommenders, we include this comparison as a sanity check: if HSTU had a large scaling advantage, it could reduce the practical relevance of transformer scaling in our setting.

We compare our medium transformer encoder to the largest HSTU configuration reported in~\citet{hstu}. Although HSTU has roughly $1.5\times$ more parameters, the transformer achieves comparable results on both pre-training metrics and ranking uplift (Table~\ref{table:scaling}). Overall, we do not observe a clear scaling-driven advantage of HSTU over a standard transformer in our setup.

\subsection{Two-Stage Training Pipeline (RQ3)}
\begin{table}[!t]
\centering
\caption{Impact of pre-training and fine-tuning stages on pair accuracy uplift}
\label{table:pretrain_finetune}
\begin{tabular}{c l c c}
\toprule
\textbf{Pre-training} & \textbf{Fine-tuning} & \textbf{Standalone} & \textbf{Feature} \\
\midrule
$\times$     & 1 year  & -7.72\% & +1.17\% \\
\checkmark   & 1 week  & -9.27\% & +0.63\% \\
\textbf{\checkmark} & \textbf{1 year}  & \textbf{-5.33\%} & \textbf{+2.32\%} \\
\bottomrule
\end{tabular}
\end{table}

We assess the effectiveness of our two-stage approach by isolating the contributions of pre-training and fine-tuning. Results in Table\,\ref{table:pretrain_finetune} show that both components are necessary for optimal performance.

Without pre-training, even a full year of fine-tuning reaches only +1.17\% uplift in the feature setting, falling short of the pre-trained + fine-tuned model (+2.32\%).

Interestingly, the fine-tuning stage must remain large even after pre-training. With the same pre-trained backbone, extending fine-tuning from one week to one year boosts feature-based uplift from +0.63\% to +2.32\% --- a substantial gain. This suggests that, unlike common LLM pipelines where the fine-tuning set is tiny relative to pre-training, impression-level ranking in our setting requires fine-tuning at a comparable scale.

Taken together, these results validate our pipeline design: pre-training provides a strong backbone, while large-scale fine-tuning aligns the model with the downstream ranking objective.

\subsection{Context Length Scaling (RQ4)}

We examine whether increasing context length (number of past historical user interactions) improves recommendation quality. We fix pre-training context length at 512 and vary only fine-tuning context length.

As shown in Table\,\ref{table:context_length}, longer histories yield consistent improvements in pairwise accuracy uplift. Increasing context length from 512 to 2048 interactions results in a notable gain (+1.01\% to +2.32\% in feature-based ranking). Extending further to 8192 interactions brings additional uplift (+2.77\%), comparable to scaling model size from 100M to 1B parameters.

\begin{table}[!t]
\centering
\caption{Effect of context length on pair accuracy uplift}
\label{table:context_length}
\begin{tabular}{c c c}
\toprule
\textbf{Context Length} & \textbf{Standalone} & \textbf{Feature} \\
\midrule
512   & -8.93\%       & +1.01\% \\
2048  & -5.33\%    & +2.32\% \\
8192  & \textbf{-4.73\%} & \textbf{+2.77\%} \\
\bottomrule
\end{tabular}
\end{table}

\subsection{Real-World Experiments (RQ5)}\label{sec:real-world}
\begin{table*}[!t]
\centering
\caption{Incremental online gains (A/B tests) from successive transformer-based models on our music streaming platform. ``TLT'' is total listening time, and ``like likelihood'' is the probability that a user presses ``like'' for a recommended item. Each row's improvement is measured relative to the previous deployments, so gains are cumulative.}
\begin{tabular}{lccrcc}
\toprule
\multirow{2}{*}{\centering \textbf{Deployment}} & 
\multirow{2}{*}{\centering \textbf{Context Length}} & \multicolumn{2}{c}{\centering \textbf{Encoder}} & 
\multirow{2}{*}{\centering \textbf{TLT}} & 
\multirow{2}{*}{\centering \textbf{Like Likelihood}} \\
\cmidrule(lr){3-4}
& & \textbf{Configuration} & \textbf{\#Params} &  & \\
\midrule
Offline V1 & 512 & L6 H512 & 18.9M & +0.52\% & +1.11\% \\
Offline V2  & 1024 & L6 H512 & 18.9M & +1.00\% & +0.73\% \\
Offline V3 & 1024 & L6 H512 & 18.9M & +0.73\% & +5.00\% \\
Real-time V1 & 1024 & L4 H256 & 3.2M & +0.32\% & +1.38\% \\
\textbf{Offline V4 (ARGUS)} & \textbf{8192} & \textbf{L10 H1024} & \textbf{126.0M} & \textbf{+2.26\%} & \textbf{+6.37\%} \\
\bottomrule
\end{tabular}
\label{table:real-world}
\end{table*}
Over the past several years, we deployed a series of transformer-based ranking models into our music recommendation pipeline. All prior models followed a two-tower architecture. They employed traditional next-item prediction pre-training (e.g., predicting the next like), followed by impression-level fine-tuning for ranking.

Table~\ref{table:real-world} summarizes the A/B test gains from each deployment:
\begin{itemize}
    \item \textbf{Offline V1}: trained on the most recent 512 engagement-based feedback events (e.g., likes)
    \item \textbf{Offline V2}: extended to consumption-based signals (e.g., streams with provided completion rate information)
    \item \textbf{Offline V3}: merged both signal types into a heterogeneous user sequence
    \item \textbf{Real-time V1}: optimized for low-latency inference by reducing encoder size $6\times$
\end{itemize}
Each new generation provided measurable uplift in total listening time (TLT) and like likelihood, evaluated via A/B tests against the existing production stack. Since all deployments were cumulative, each improvement built upon prior transformer models and other non-deep-learning enhancements.

\textbf{Offline V4 (ARGUS)} combines several advancements: a significantly larger encoder (126M parameters), extended user context (up to 8192 events), and the proposed training pipeline described in this work. ARGUS achieved the strongest transformer-driven online gains to date: +2.26\% in total listening time and +6.37\% in like likelihood. While measuring online uplift for the largest 1B configuration would be ideal, it was out of scope due to the high computational cost of large-scale online experiments. 

Since the initial submission of this work, we have further integrated a smaller real-time variant of ARGUS directly into the production ranker, observing complementary gains when combined with the daily offline embeddings. 

\section{Discussion}
\label{sec:discussion}

\subsection{Relation to HSTU}
\label{sec:discussion-hstu}
In addition to the encoder comparison in Section~\ref{sec:rq2-hstu}, we highlight several practical differences between HSTU~\cite{hstu} and ARGUS that are important for interpreting results.

\paragraph{Two pipelines vs. a unified formulation.}
The original HSTU paper employs two distinct procedures: a ranking task with
interleaved item-action tokens and a retrieval task that incorporates
negative interactions but limits predictions to positive events. In contrast, ARGUS utilizes a unified pre-training objective, combining feedback prediction (conceptually similar to HSTU’s ranking loss) with a second task that predicts all item interactions, including negative ones.

\paragraph{Context modeling.}
HSTU represents context changes as separate events. ARGUS treats context as part of each interaction: either as an explicit token in the $(c_t,i_t,f_t)$ stream (full version) or fused into a single interaction embedding (simplified version).

\paragraph{Serving cost for ranking.}
ARGUS is designed to be inexpensive at serving time, especially in the offline setting: we fine-tune into a two-tower model with daily user embedding refresh and dot-product scoring (Section~\ref{sec:fine-tuning}). This makes impression-level ranking substantially cheaper than real-time target-aware schemes that require encoding the user state jointly with each candidate item.

\subsection{Practical Considerations}
\label{sec:practical}

\paragraph{Generalization across domains.}
Although this paper focuses on music recommendation, we have applied the same architecture in other large-scale domains, including an e-commerce marketplace, grocery delivery, and advertising, observing consistent gains.

\paragraph{Cold start.}
ARGUS is not tied to an itemID lookup: item representations can be formed from arbitrary content and metadata features, and the ItemID feature can be omitted altogether. In our deployments across multiple domains, we experimented with different item-feature configurations and observed consistent performance, suggesting that the approach does not have fundamental cold-start limitations.

\paragraph{Additional fine-tuning scenarios.}
Since the initial submission, we have explored additional fine-tuning scenarios beyond impression-level ranking. In particular, in several domains we have already deployed a candidate-generation model: using the same pre-training recipe, we fine-tune the backbone to predict future positive interactions. We also tested more challenging objectives, such as generating the user’s next session as a whole, and found that the same backbone can support these richer targets.

\section{Conclusion}
We present a scalable framework for training large recommender transformers, successfully deployed in a real-world music recommendation system. Drawing inspiration from reinforcement learning and advances in large language models, we introduce a novel autoregressive pre-training task that unifies next-item and feedback prediction, encouraging the model both to imitate observed behavior and generalize beyond it.

\begin{acks}
We would like to thank the Yandex Music Recommendations team, led by Daniil Burlakov, for their continuous support throughout this work. Over the years, they have generously shared their expertise in the music domain and have provided invaluable assistance with deployments, including the ARGUS deployment. It has been a real pleasure collaborating with them.
\end{acks}



\bibliographystyle{ACM-Reference-Format}
\balance
\bibliography{sample-base}

\appendix
\end{document}